\documentclass[a4paper,11pt]{article}
\pdfoutput=1
\usepackage{jinstpub}
\usepackage{lineno}
\usepackage{caption}
\usepackage{subcaption}

\title{Upgrading gSeaGen: from MeV to PeV neutrinos}

\author[a,b,1]{A. Garcia Soto,\note{Corresponding author.}}
\author[c]{C. Distefano,}
\author[d]{P. Kalaczyński}

\affiliation[a]{Instituto de Física Corpuscular (IFIC), Consejo Superior de Investigaciones Científicas (CSIC) y Universitat de València (UV), 46980 Paterna, València, Spain.}
\affiliation[b]{Harvard University, Dept. of Physics, Cambridge, MA 02138, USA}
\affiliation[c]{INFN, Laboratori Nazionali del Sud, Via S. Sofia 62, Catania, 95123, Italy}
\affiliation[d]{National Centre for Nuclear Research,\\Pasteura 7, Warsaw, Poland}

\emailAdd{aagarciasoto@km3net.de}

\abstract{The gSeaGen framework has been upgraded to simulate events detectable in neutrino telescopes induced by neutrino or cosmic ray interactions. The new version is well-suited to generate neutrino interactions at energies from a few MeV to EeV, profiting from the latest GENIE extensions to lower and higher energies. In addition, a brand new functionality to propagate leptons from CORSIKA air showers has been developed. Novel features of gSeaGen will be presented, related to simulation of PeV tau neutrinos and charged lepton propagation, using the KM3NeT detectors as physics case.}


\collaboration[c]{on behalf of KM3NeT collaboration}

\proceeding{Very Large Volume Neutrino Telescope Conference\\
18-21 of May 2021\\
Valencia (Online)}

\begin{document}
\maketitle
\flushbottom

\section{Introduction}

Under-water/ice Cherenkov neutrino detectors, also known as neutrino telescopes, have a unique feature as they can detect neutrinos in a vast energy range between few MeV and PeV. In particular, the KM3NeT collaboration is building two different detectors (ORCA and ARCA) that will be sensitive to neutrinos of MeV, GeV and PeV energies \cite{km3net}. Hence, one of the most challenging tasks for these experiments is to understand distinct aspects (e.g. neutrino sources, transport through the planet and generation of events near the sensitive volume) in very different energy regimes.  

The gSeaGen framework \cite{gseagen} was designed to efficiently simulate neutrino interactions in large volumes accounting for the aforementioned points, and for this reason is a very suitable generator for neutrino telescopes. In its early stages, the package was mainly developed for KM3NeT. Nevertheless, the current version of the code is open source and can be easily customized to the needs of other neutrino telescopes.

\begin{figure}[htbp]
\centering 
\includegraphics[width=0.83\textwidth]{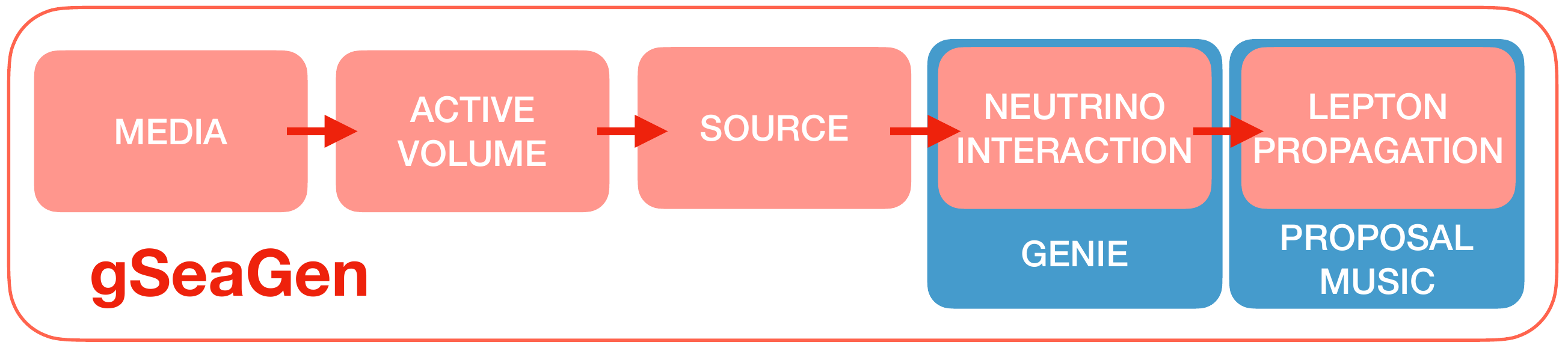}
\caption{Block diagram of gSeaGen. Blocks show the main steps that are processed in the package and arrows illustrate the sequence to generate an event in the detector.}
\label{fig:workflow}
\end{figure}

Fig.~\ref{fig:workflow} shows a diagram illustrating the main steps used to generate events with gSeaGen. One key feature of the code is that it allows the user to change the properties of the media (e.g. ice, water, rock) as well as detector locations (i.e. depth and geographic coordinates). In addition, the code includes a driver to automatically calculate the volume in which an interaction can produce detectable particles depending on the energy and flavor of the incoming neutrino. It is also able to generate neutrinos coming from diffuse, point-like and extended sources (both steady and transient). Finally, the code relies on external packages like GENIE \cite{genie}, MUSIC \cite{music} or PROPOSAL \cite{proposal} to simulate the neutrino interactions and the propagation of charged leptons.   

In this work we describe new features that are available in the code, particularly the extension to the high energy regime profiting from recent developments in GENIE \cite{genie_v3}. The generation of neutrino events above 1 TeV in a KM3NeT ARCA-like detector geometry is discussed in Sec.~\ref{sec:neutrino}. A comparison between different muon propagation codes is done in Sec.~\ref{sec:propagation}.

\section{Neutrino interaction}
\label{sec:neutrino}

The previous version of gSeaGen \cite{gseagen} was limited to generating neutrinos up to 5 TeV due to GENIE's range of validity. The newest version of GENIE includes new modules (called HEDIS and GLRES) that compute neutrino differential cross sections up to 1 EeV. This new version has been tested in gSeaGen and results are presented in the following paragraphs.

In addition to the extension to generate neutrino interactions of any flavor at high energies, the new version simulate tau leptons differently. In the first version of gSeaGen, the outgoing tau from Charged Current interactions was decayed on the spot and PYTHIA6 \cite{pythia} was used to compute the kinematics of the decay products. This method is a good approximation at energies below 1 TeV but is not valid at higher energies, hence several aspects for the generation of tau neutrinos have been changed. First, the calculation of the interaction volume is done accounting for the decay length and products of the outgoing lepton. Second, the propagation of the tau includes energy losses using TAUSIC \cite{music}. Finally, the kinematics of the decay products are computed using TAUOLA \cite{tauola}.  

\begin{figure}[htbp]
\centering 
\includegraphics[width=1\textwidth]{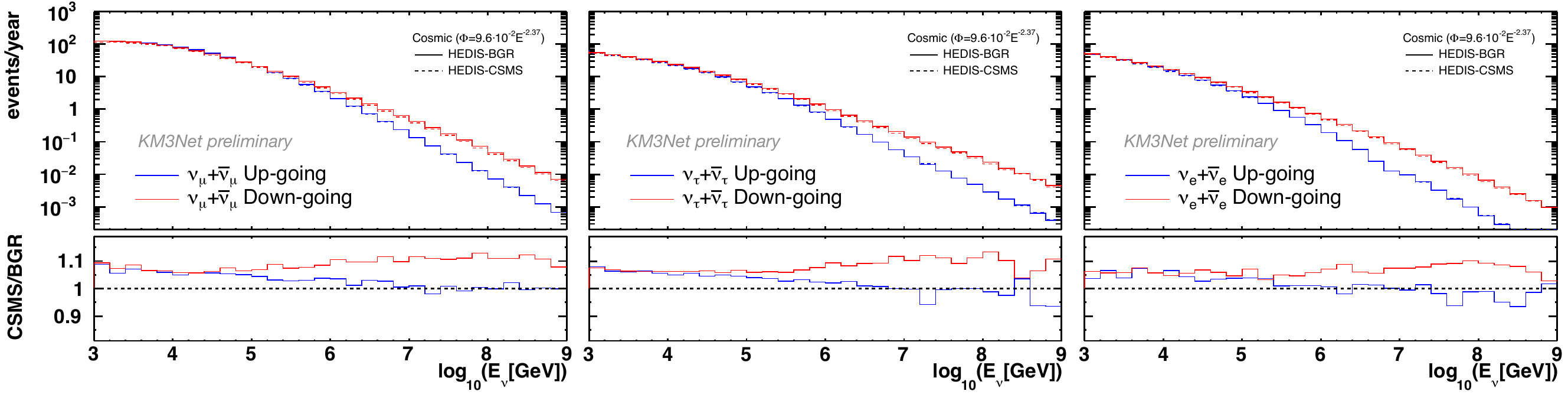}
\caption{Muon (left), tau (center), electron (right) neutrino and anti-neutrino energy distributions generated with gSeaGen for up- and down-going neutrinos. The diffuse flux shown in the legend is assumed. Two different cross section models are compared: CSMS (dashed lines) and BGR (continuous lines).}
\label{fig:neutrinos}
\end{figure}

Fig.~\ref{fig:neutrinos} shows the energy spectrum for the three neutrino species producing detectable events in one building block of ARCA assuming a diffuse flux. Two different DIS models implemented in HEDIS (CSMS and BGR) are compared. The rate of down-going events is mainly dominated by the total neutrino-nucleus cross sections, whereas up-going events are affected by Earth absorption at high energies. The muon neutrino contribution dominates at these energies as muons originated in CC interactions travel long distances and reach the active volume of the detector. Above 10 TeV the tau and electron neutrino rates differ since the tau leptons can propagate as muons before they decay.

\begin{figure}[htbp]
\centering 
\includegraphics[width=0.8\textwidth]{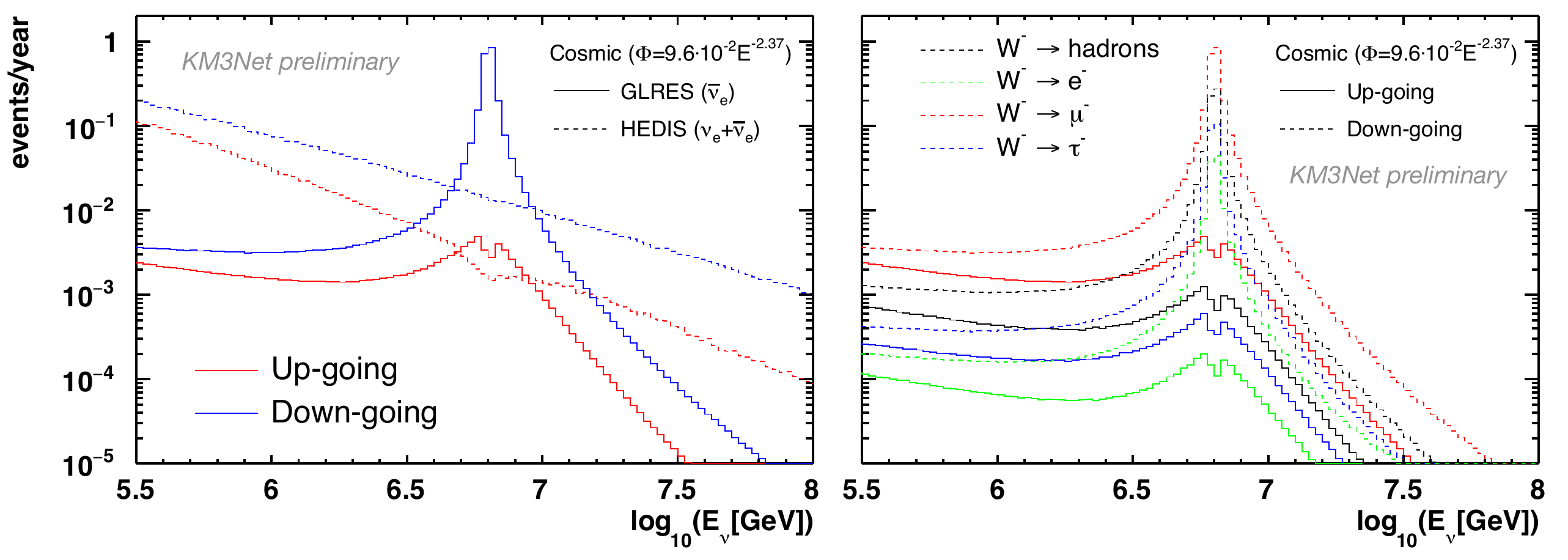}
\caption{Electron anti-neutrino energy distributions generated with gSeaGen for up- and down-going neutrinos interacting via the Glashow resonance. Left: comparison with the predictions for DIS interactions for electron neutrinos and anti-neutrinos. Right: breakdown in the different decay channels of the W boson.}
\label{fig:glashow}
\end{figure}

Another important development is related to the generation of Glashow resonant events \cite{glashow} since the latest version of GENIE allows to simulate this type of interactions as well. Most of the studies of this channel have been focused on trying to observe fully or partially contained showers in the detector coming from the decay of a W boson into hadrons \cite{glashow_ice}. To simulate these events the interaction volume can be restricted to the convex hull of the detector as hadrons will travel only a few meters. However, when the W boson decays into a muon or a tau the interaction volume must be significantly larger. gSeaGen can compute different interaction volumes depending on the decay channel. Fig.~\ref{fig:glashow} shows the expected rate of Glashow resonant events in one ARCA block. One can observe that the main contribution comes from muons even though its branching ratio is smaller than the hadronic decay.
 
\section{Muon propagation}
\label{sec:propagation}

A fundamental step in the simulation of events around the detector volume is the propagation of muons. On the one hand, muons produced in neutrino interactions can travel long distances and produce light in the detector. On the other hand, muon bundles from extensive air showers can also be detected at large depths. Hence, it is crucial to properly simulate the energy losses and scattering as muons propagate through matter. Currently, gSeaGen includes two internal implementations to propagate muons in water/ice and rock called PropaMuon and JPP. Additionally, the code can be linked to two external open source programs: MUSIC \cite{music} and PROPOSAL \cite{proposal}. 

\begin{figure}[htbp]
\centering
\includegraphics[width=0.45\textwidth]{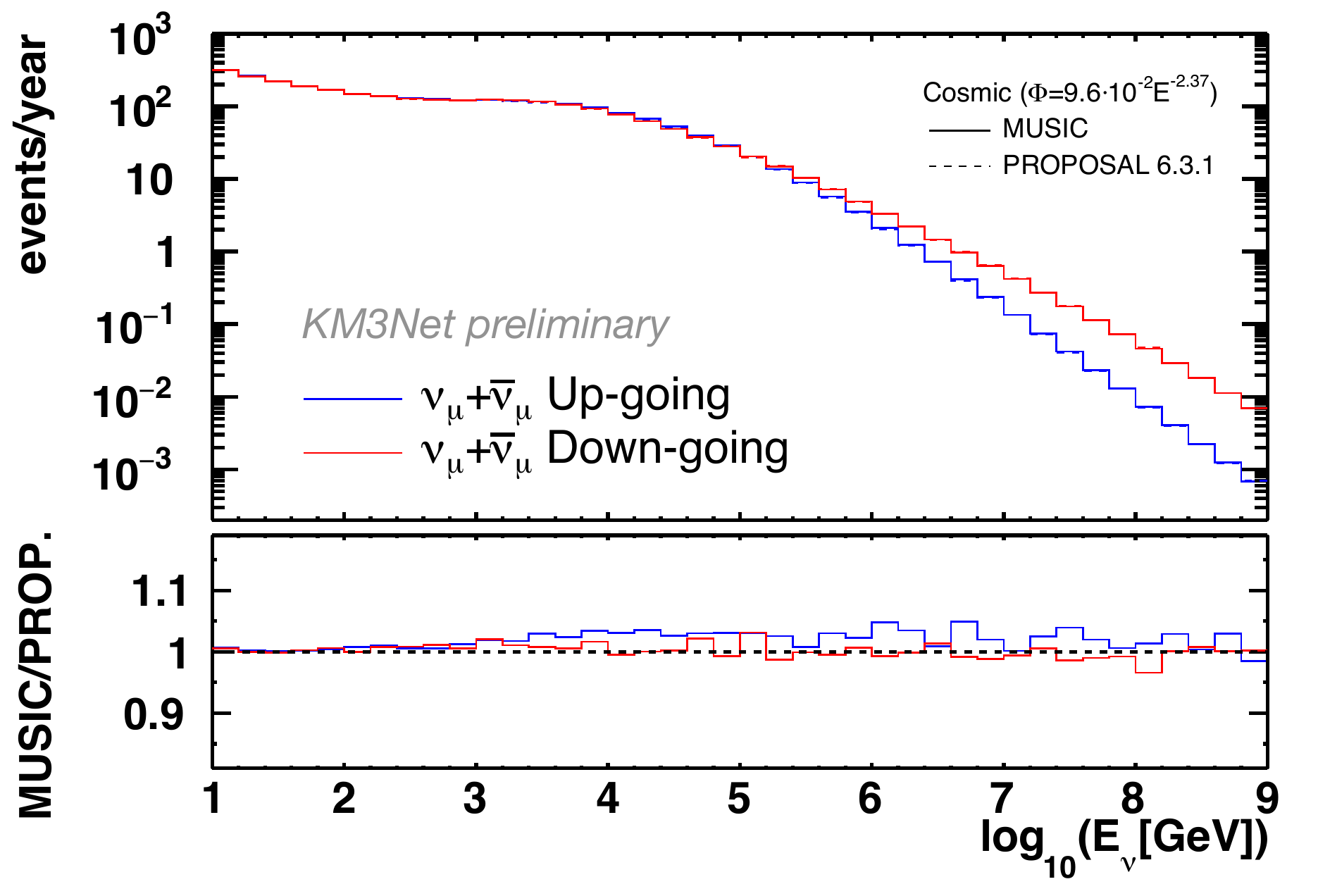}
\includegraphics[width=0.42\textwidth]{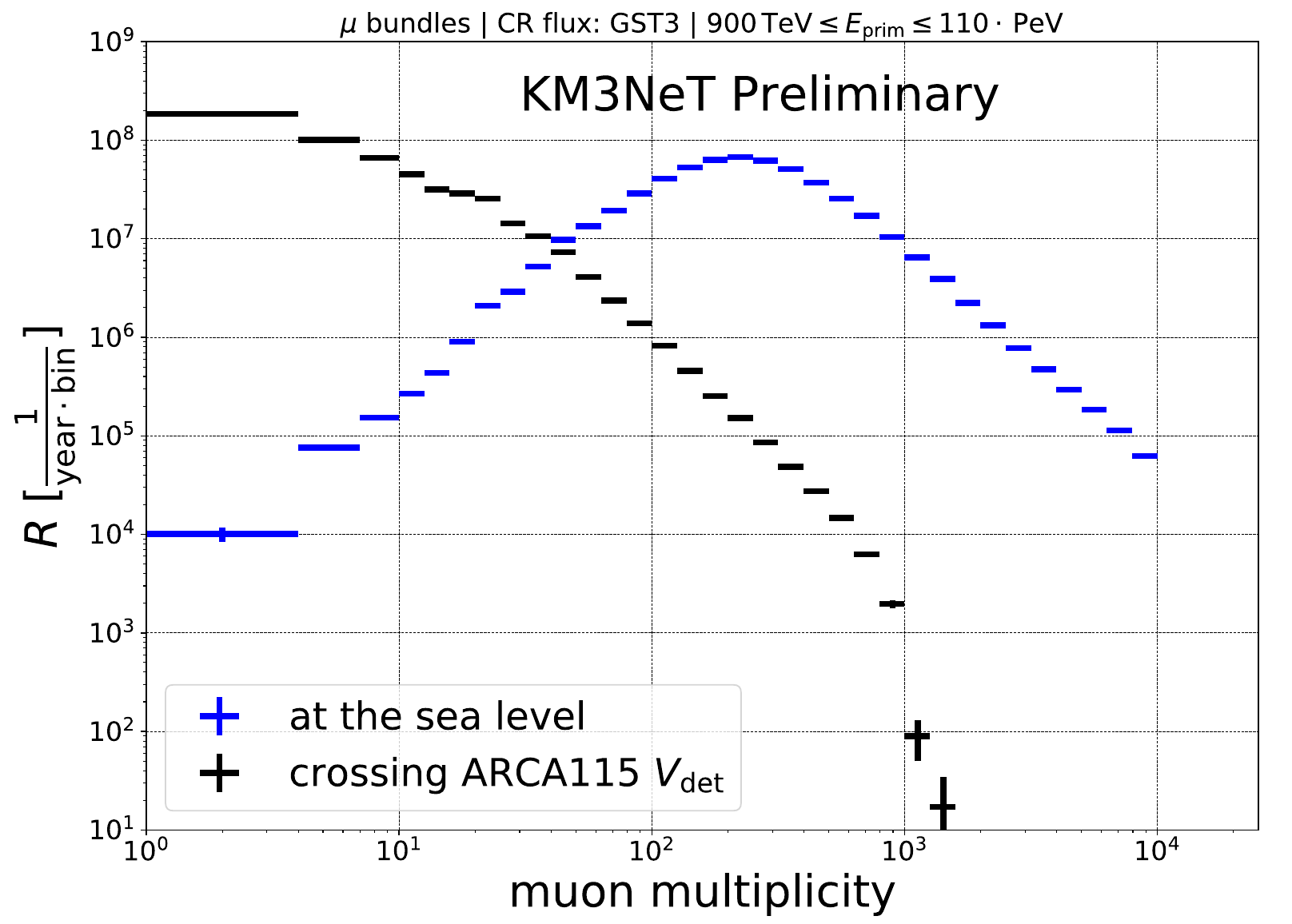}
\caption{Left: Muon (anti-)neutrino energy distributions generated with gSeaGen for up- and down-going neutrinos. The results obtained with MUSIC and PROPOSAL are compared and found to be consistent between each other. Right: Comparison of muon multiplicity distributions at the sea level and after propagating to the ARCA-like detector geometry. The propagation is done using PROPOSAL on atmospheric muons simulated with CORSIKA.}
\label{fig:lepton}
\end{figure}

Fig.~\ref{fig:lepton} shows the energy spectrum for the muon neutrinos producing detectable events in one building block of ARCA assuming a diffuse flux. Two different muon propagation codes (MUSIC and PROPOSAL) are compared for up-going and down-going events (i.e. rock and water propagation respectively). The agreement between the two codes is better than 5\% and only subtle differences are observed when propagating muons through rock. Recently, direct processing of CORSIKA output files \cite{corsika} has been implemented in gSeaGen. Consequently, it is possible to propagate muons from the air showers until any custom depth of ice or water as shown in Figure \ref{fig:lepton}.

\section{Conclusions}

This work details the main developments of gSeaGen with respect to the previous version. The two main new features are: possibility of generating neutrino events up to PeV energies using the latest version of GENIE, and a new tool that allows processing CORSIKA output files to propagate air showers in water or ice. This upgrade is already publicly available\footnote{https://zenodo.org/record/4766015\#.YKOtspMzau4}.

\acknowledgments

A.G. acknowledges support from the European Union’s H2020-MSCA Grant Agreement No. 101025085.

\end{document}